\documentclass[sigconf]{acmart}

\usepackage{booktabs} 
\usepackage[utf8]{inputenc} 
\usepackage{balance}

\setcopyright{none}

\renewcommand\footnotetextcopyrightpermission[1]{} 
\author{Younghun Song}
\affiliation{%
  \institution{KAIST}
  \streetaddress{291 Daehak-ro, Yuseong-gu}
  \city{Daejeon}
  \country{Korea, Republic of}
  \postcode{34141}
}
\email{younghun.song@kaist.ac.kr}

\author{Jae-Gil Lee}
\authornote{Corresponding Author}
\affiliation{%
  \institution{KAIST}
  \streetaddress{291 Daehak-ro, Yuseong-gu}
  \city{Daejeon}
  \country{Korea, Republic of}
  \postcode{34141}
}
\email{jaegil@kaist.ac.kr}

\begin{document}

\title{Augmenting Recurrent Neural Networks with High-Order User-Contextual Preference for Session-Based Recommendation}

\begin{abstract}
The recent adoption of recurrent neural networks (RNNs) for session modeling has yielded substantial performance gains compared to previous approaches. In terms of context-aware session modeling, however, the existing RNN-based models are limited in that they are \emph{not} designed to explicitly model rich static user-side contexts (e.g., age, gender, location). Therefore, in this paper, we explore the utility of explicit user-side context modeling for RNN session models. Specifically, we propose an \emph{augmented RNN (ARNN)} model that extracts \emph{high-order user-contextual preference} using the product-based neural network (PNN) in order to augment any \textit{existing} RNN session model. Evaluation results show that our proposed model outperforms the baseline RNN session model by a large margin when rich user-side contexts are available.
\end{abstract}

%
%



\maketitle

\section{Introduction}

Making recommendations based on session logs of user-item interactions has been a major challenge for the recommender systems community. Utilizing session logs for recommendations has obvious advantages: 1) they can be used to infer user preferences to make recommendations to anonymous/fresh users, and 2) they can provide a more personalized recommendation that matches a user's current interest.

Since the pioneering work of Hidasi et al.\cite{gru4rec}, recurrent neural networks (RNNs) have been the de-facto choice for modeling session-based recommender systems, as they are capable of effectively exploiting the sequential nature of user session data. Inspired by this success, numerous RNN-based session models have been proposed to accommodate diverse aspects of session-based recommender systems---e.g., by incorporating item content features\cite{parallel-gru4rec}, modeling latent user intent\cite{ed-rec}, or merging information from past and current user sessions\cite{ii-rnn, hgru4rec}.

Another interesting application of RNNs has been their usage in context-aware sequential recommendations\cite{crnn,ca-rnn}, which can be also applied to session-based recommendation settings. However, a majority of the research in this line focused on exploiting \emph{dynamic} temporal contexts(e.g., time-of-the-day), and there have been few attempts that explicitly consider the \textit{static} user-side contexts(e.g., age, gender, location), which are readily available in commercial systems. In terms of session modeling, explicit modeling of static user contexts is important because \emph{user navigation paths} of items during sessions are dependent on static contexts, due to the fact that users with different static contexts have different preferences with respect to the same items. For example, when recommended a job position in Paris during a session, a user living in New York will tend to look for alternative positions in New York in the following searches, while a user living in Paris will tend to look for the similar positions.

Therefore, in order to seamlessly integrate static user-side contexts into RNN session models, we propose an \emph{augmented RNN (ARNN)}, which is an augmented version of RNN that can improve on any \textit{existing} RNN session model. The unique feature of ARNN is that it estimates \emph{user-contextual preference} by modeling high-order interaction between static user context and the previous item using a product-based neural network (PNN)\cite{product-based-nn}, which is a neural network (NN) variant of the factorization machine (FM)\cite{factorization-machines}. By integrating this contextual preference with the hidden states of an RNN session model, the ARNN makes more personalized recommendations than the plain RNN session models that does not consider user contexts.

We evaluated the ARNN on two datasets: a job recommendation dataset and a standard e-commerce dataset. Through an experiment, we verified that the ARNN successfully captured user-contextual preference when abundant user contexts were available, which led to a substantial improvement in the recommendation quality against an RNN baseline.

\section{Related Works}
\subsection{Session-based / Context-aware Recommendation with RNNs}\label{sess-rec-with-rnns}
Traditional session-based recommendation algorithms have largely been based on item-to-item approaches\cite{item-based-cf-sarwar,item-based-cf-linden}. However, the recent success of deep learning has led to the adoption of recurrent neural networks (RNNs) to session-based recommendations, starting from GRU4REC\cite{gru4rec}. The motivation of GRU4REC was to mitigate the consistently-cold-starting problem that occurs frequently in a typical e-commerce environment, where most of the customers do not log in and thus are anonymous. After GRU4REC, several RNN-based models were introduced to consider various aspects of session-based recommendations \cite{ed-rec, ii-rnn, hgru4rec, parallel-gru4rec}.

RNNs also started to gain momentum in the field of context-aware sequential modeling. Some approaches used temporal contexts like time-of-the-day or time difference between the previous and current interaction\cite{crnn, ca-rnn}. There was also an attempt to use the type of interaction events as a contextual information\cite{crnn}. However, all those contexts were \emph{dynamic} contexts dependent on specific transaction instances, and no models have been introduced that can equip the powerful RNN session models with abundant static user-side contexts like age, location, or current job title.\footnote{For convenience, we'll use the term ``user context'' to refer to  ``static user context'' throughout this paper.}

\subsection{PNN and FM}

In terms of context-aware recommendations, the factorization machine (FM)\cite{factorization-machines} and its neural network (NN) variants\cite{product-based-nn,nfm} have been widely successful, particularly in applications such as click-through-rate (CTR) prediction. The FM and its variants build the latent embeddings for the categorical contexts in order to estimate the 2nd-order interactions between categorical features, thereby aiming to resolve the data sparsity inherent in recommender systems. Among the NN variants of FM, the product-based neural network (PNN)\cite{product-based-nn} has been known to effectively capture the \textit{higher-order interactions} between the categorical variables by applying pairwise product operations for each possible pair of embedded categorical fields and passing the resulting vector as the input to a NN.

\section{Model Description}

\subsection{Problem Setup}\label{problem-setup}
The goal of our work is to capture the contribution of user context information in determining a user's \textit{item navigation path} during an online session. Therefore, our dataset can be expressed as follows:
\begin{itemize}
    \item Session: $s_i=\{(x_t, y_t)\}_{t=1}^{T_i}$, where $y_t$ is the item that the user interacted with at $t$ within the session, and $x_t$ is the user \textit{contextual vector}.
    \item Training set: $\mathcal{D}_{train}=\{s_1,\dots,s_{|\mathcal{D}_{train}|}\}$
    \item Test set: $\mathcal{D}_{test}=\{s'_1,\dots,s'_{|\mathcal{D}_{test}|}\}$
\end{itemize}
Notice that $x_t$ is included in every user session. $x_t$, a user contextual vector, is a concatenation of the one-hot encodings for each categorical fields, as in Rendle\cite{factorization-machines}. For instance, if we have the context information \texttt{(Gender=Female,Location=U.S.)} about a user, then each field are first encoded separately as $[1,0], [0,0,1,0,\dots,0]$ and concatenated to produce the final input $x_t=[1,0,0,0,1,0,\dots,0]$. The objective of the ARNN is, given a new sequence of (context,item) pairs in a user session $s'_u=\{(x_1,y_1),\dots,(x_{t-1},y_{t-1})\}$ included in a test set, to predict the next item $y_{t}$ with which the user will interact.

Given the dataset, our problem reduces to a sequential \textit{ranking} task with \textit{implicit feedback}, where the input consists of the item indices $y_t$'s and user contexts $x_t$'s, and the output consists of the ranking scores for the positive items $r(y_t)$ and negative items $r(y_n)$. Now, following the approach by Rendle et al.\cite{bpr} and introducing the time-independence assumption for user contexts, we can set up the following joint likelihood for a single session as the objective function:
\begin{equation}
    \mathcal{L}=\prod_{t=1}^T\prod_{n=1}^N{p(r(y_t)>r(y_n)|\mathbf{y}_{<t},x_{t-1})}
    \label{eqn:likelihood}
\end{equation}
where $\mathbf{y}_{<t}=(y_{t-1},\dots,y_1)$.

\subsection{Model Overview}\label{model-overview}
The proposed model, which we call the \textbf{augmented RNN (ARNN)} model, augments an RNN session model with a PNN \textit{context encoder}. At the training stage, a PNN context encoder and an RNN session model are first pretrained separately to optimize the likelihood \eqref{eqn:likelihood}, where the negative items are defined using the items in the session-parallel mini-batches as in GRU4REC\cite{gru4rec}. Then, a \textit{merging layer} is trained on top of the pretrained PNN and RNN to integrate the information from both networks. After training, the ARNN scores the items conditioning on both the PNN-produced contextual preference $c_t$ and the RNN hidden state $h_t$. The details of the PNN/RNN components and the merging network are explained in the following sections.

\subsection{PNN Context Encoder}\label{pnn-context-encoder}
\begin{figure}[h!]
    \centering
    \includegraphics[width=0.5\textwidth, angle=0]{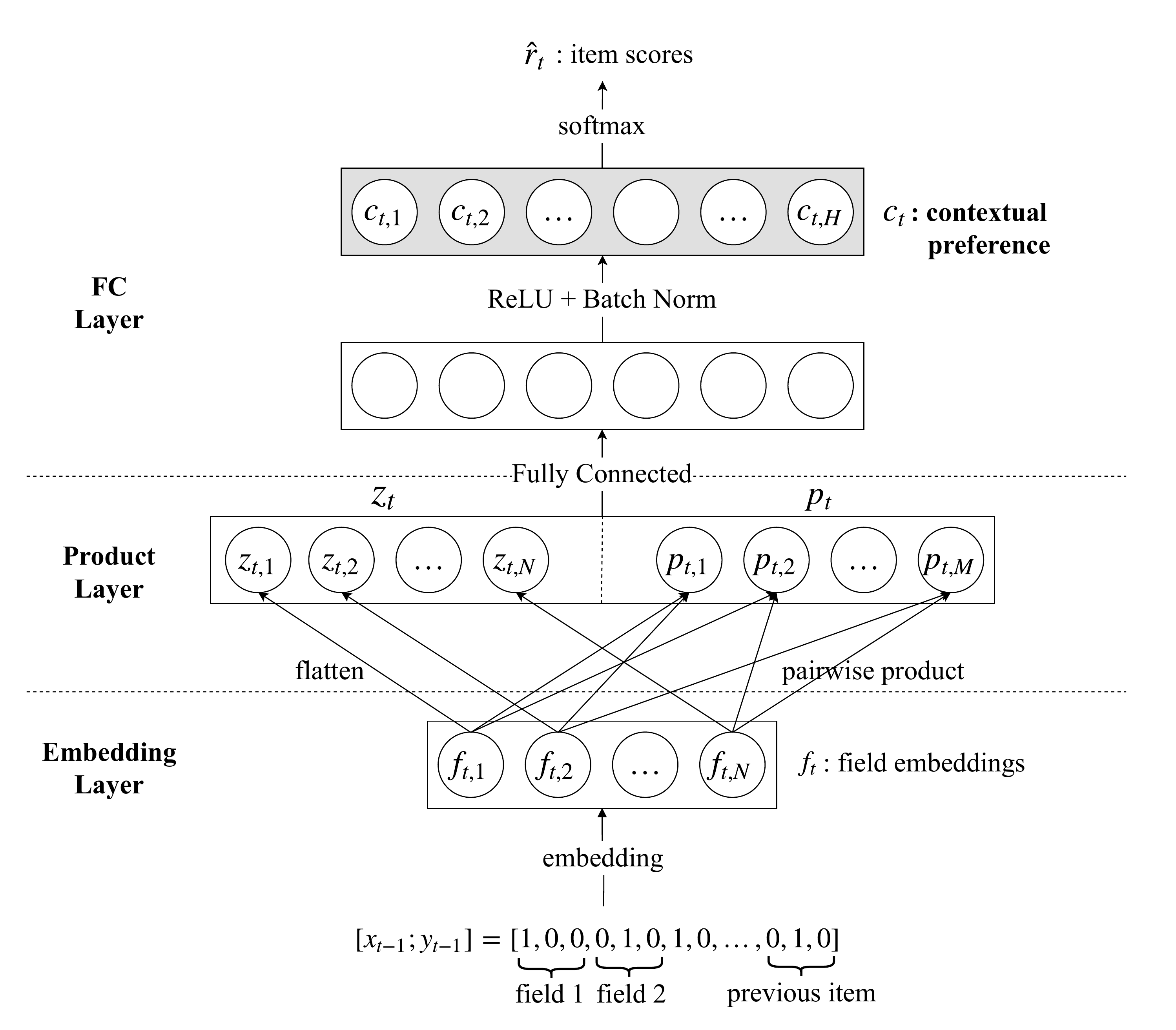}
    \vspace*{-0.3cm}
    \caption{Architecture of the PNN context encoder.}
    \label{fig:IPNN}
\end{figure}

The foundational element of the ARNN is the PNN context encoder that models user-contextual preference by capturing high-order interactions between user contexts and previous items. Similar to the original PNN\cite{product-based-nn}, the PNN context encoder is composed of three layers: embedding layer, product layer, and FC layer, as shown in Figure \ref{fig:IPNN}.

First, $x_{t-1}$ (as explained in Section \ref{problem-setup}) and the previous item $y_{t-1}$ are \textbf{concatenated} and enters the embedding layer, producing the embeddings for each fields(including the previous item). As the original PNN, only one position remains active per field.

Next, pairwise inner product operations are applied to each possible pair of field embeddings to produce the pairwise signal $p_t$. At the same time, all field embeddings are flattened and concatenated to produce the linear signal $z_t$.

Then, $p_t$ and $z_t$ are concatenated and fed as an input to the FC layer, followed by a rectified linear unit (ReLU) + batch normalization \cite{batch-norm}. From this nonlinear transformation of both linear/pairwise signals, the PNN can effectively model high-order interactions between the user contexts and the previous item, obtaining a vector encoding the user-contextual preference.

Finally, after the batch normalization layer, the user-contextual preference $c_t$ goes through the softmax layer that calculates the final item scores. The item scores are then used to calculate the ranking loss and to pretrain the PNN, as mentioned in Section \ref{model-overview}.

\subsection{RNN Session Model} \label{rnn-session-model}
For simplicity, RNN session models are set to follow the simple gated recurrent unit (GRU)\cite{rnn-encoder-decoder} architecture proposed in GRU4REC\cite{gru4rec}. More specifically, let $y_t$ be the one-hot encoded item for the $t$-th transaction and $h_t$ be the corresponding GRU hidden state. The scores for the next items are then calculated by the GRU as follows:
\begin{align*}
    \vspace*{-0.3cm}
    &\hat{r}(y_{t}|\mathbf{y}_{<t})=softmax(h_t) \quad h_t=GRU(y_{t-1},h_{t-1})
\end{align*}
Similar to Section \ref{pnn-context-encoder}, the item scores are then used to pretrain the GRU for the session-based ranking task. However, our model does not restrict the RNN session model to take the proposed simple GRU form, so \emph{any} existing RNN session models can be used instead of the GRU.

\subsection{Augmenting the RNN with the PNN}
\begin{figure}[h!]
    \centering
    \includegraphics[width=0.35\textwidth, angle=0]{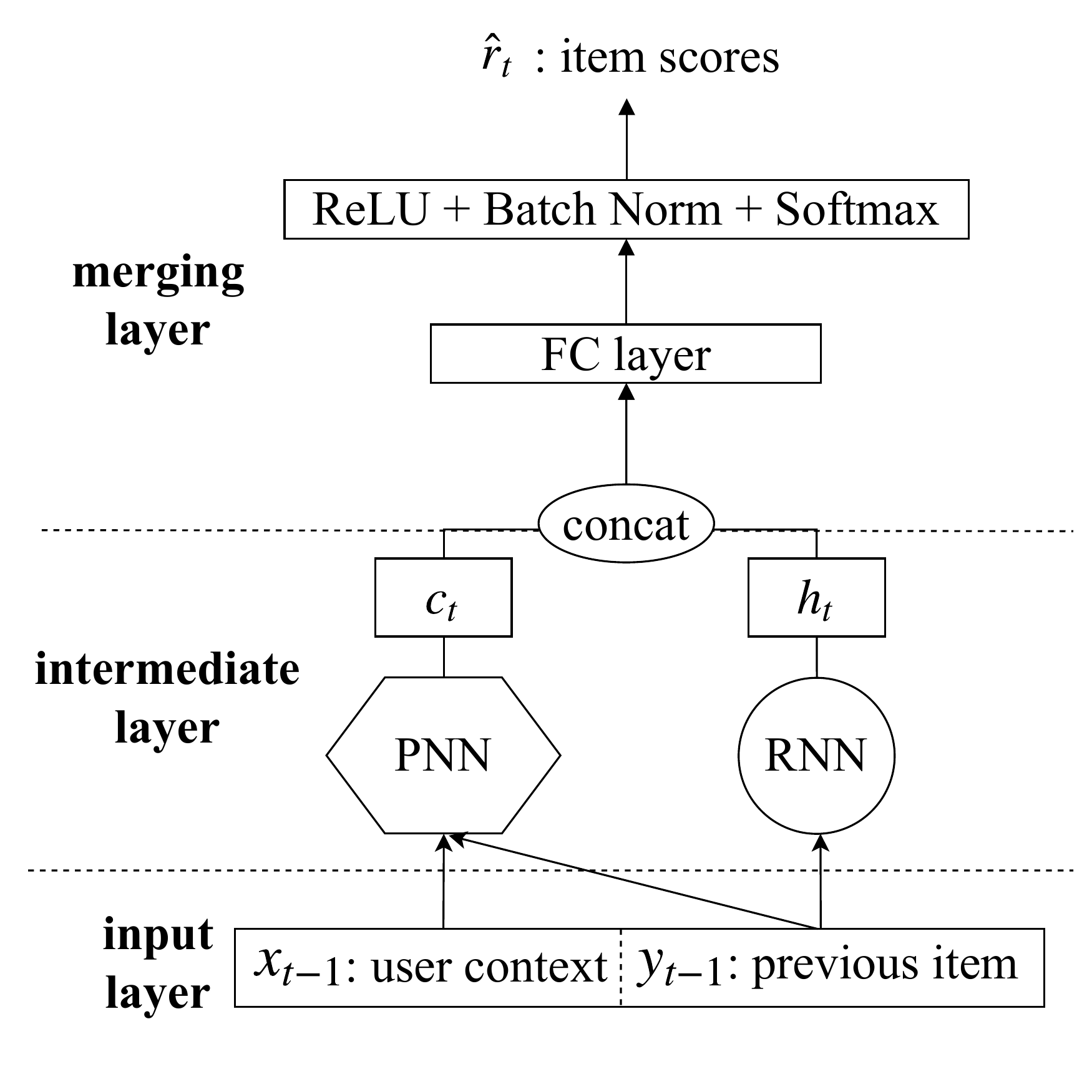}
    \vspace*{-0.3cm}
    \caption{Overall architecture of the ARNN model.}
    \label{fig:ARNN}
\end{figure}
Figure \ref{fig:ARNN} shows the overall architecture of the ARNN which consists of three types of layers: the input layer, the intermediate layer, and the merging layer. The input layer is simply where the model receives its input data. The intermediate layer consists of the PNN/RNN that we pretrained(as decribed in Sections \ref{pnn-context-encoder} and \ref{rnn-session-model}). To use the pretrained PNN/RNN as \emph{feature extractors}, we remove the final softmax layers from both the PNN/RNN, and freeze the parameters so that they are not further updated\footnote{However, we found that retraining only the batch normalization layers for the PNN slightly increases the model performance.}. The \textit{merging layer} $M$ is where the PNN output $c_t$ and the RNN hidden state $h_t$ are combined and used to calculate the item scores as follows:
\begin{align*}
    \vspace*{-0.3cm}
    &\hat{r}(y_{t}|\mathbf{y}_{< t},x_{t-1})=softmax(M([c_t;h_t]))\\
    &c_t=PNN(x_{t-1},y_{t-1}) \quad h_t=GRU(y_{t-1},h_{t-1})
\end{align*}
In our model, $M$ is a simple FC layer followed by a ReLU + batch normalization layer.



\section{Experiments}

\subsection{Dataset}
To test our algorithm, we used two datasets: the XING dataset from the RecSys 2016 Challenge and the TMALL dataset from the IJCAI-15 competition. Table \ref{tab:dataset} presents the profiles of the two datasets. The XING dataset contains the user-item interaction logs from \texttt{xing.com}, a social networking service specialized for job searching. Rich user-side categorical contexts such as job roles and career levels are included in the XING dataset. Among those contexts, we picked only 12 attributes (e.g.,  job roles, career level, and country/region) that are likely to be useful for predicting job preferences. By contrast, the TMALL dataset is a standard e-commerce dataset in which only three types of user-side categorical information (user ID, age, and gender) are available. We thus used all three attributes. The purpose of employing the TMALL dataset was to examine how the degree of abundance in user-side contexts impact the performance of the proposed context-augmented model.
\begin{table}[h!]
\centering
    \caption{Dataset statistics.}
    \vspace*{-0.3cm}
      \label{tab:dataset}
      \begin{tabular}{lcc}
        \toprule
        & XING & TMALL \\
        \cmidrule{1-3}
        \# of users & 769K & 424K \\
        \# of items & 1M & 1.1M \\
        \# of sessions & 2.1M & 6.5M \\
        \# of transactions & 7.2M & 48.6M \\
        \# of user-side categorical contexts used & 12 & 3 \\
        \bottomrule
      \end{tabular}
    \vspace*{-0.3cm}
\end{table}

\subsection{Preprocessing}
\subsubsection{Train/test split}
For both datasets, we extracted the last three days for testing purposes and trained on the preceding 27 days. Transactions including the items not in the training set were ignored as in Hidasi et al.\cite{gru4rec}.
\subsubsection{Session ID marking}
Since session IDs were missing in both XING and TMALL, we manually set time thresholds for marking a sequence of transactions as a session. The threshold was 1 hour for XING, and 1 day for TMALL (as in Jannach et al.\cite{sess-rnn-meets-neighborhood}).
\subsubsection{Sampling on the items}\label{sampling-on-the-items}
Due to scalability issues, only the top items that covered 50\% of the transactions were selected for modeling. In other words, the items were sorted by popularity and a minimum item popularity threshold was set such that the transactions including only the items above the threshold accounted for 50\% of the entire dataset. After this sampling, we were left with 13,259 items for XING and 11,939 items for TMALL.
\subsubsection{Encoding the categorical attributes}
All categorical features were encoded as binary features that indicated the presence of the corresponding features. For XING, there were multi-valued categorical features that held numerous values (e.g., \texttt{jobroles=[350, 891]}). Making binary features for all the multi-valued features resulted in prohibitively high-dimensional input to the PNN, which incurred large computational cost. Thus, similar to the process we described in Section \ref{sampling-on-the-items}, we created binary features for only the most popular categories that covered 75\% of the transactions. The least-popular 25\% of the features were mapped to a single "unknown" attribute. After this encoding, the total number of input fields required for the PNN was 368 for XING and 4 for TMALL, including the field kept for embedding the previous items.

\subsection{Optimization and Hyperparameters}
\subsubsection{Loss Function}
The TOP1 loss\cite{gru4rec} was used as a ranking loss to maximize the objective \eqref{eqn:likelihood}. Let $N_s$ be the number of negative samples in a session-parallel mini-batch. In addition, let $\hat{r}_{s,i}$ be the calculated score for the $i$-th item in the session $s$. Then, the TOP1 loss is defined as:
\begin{equation*}
    TOP1_{s,i} = \frac{1}{N_s}\sum_{j=1}^{N_s}{\left(\sigma(\hat{r}_{s,j}-\hat{r}_{s,i})+\sigma(\hat{r}_{s,j}^2)\right)}
\end{equation*}
This loss function is intended to push the scores of the negative items to zero, which prevents the overall item scores from exploding during optimization.

\subsubsection{Hyperparameters and Optimizer}
Here, we provide information about the core hyperparameters and optimizer. The hyperparameters for the GRU4REC follows that of Quadrana et al.\cite{hgru4rec} for XING, and Jannach et al.\cite{sess-rnn-meets-neighborhood} for TMALL.
\begin{itemize}
    \item GRU4REC
    \begin{itemize}
        \item GRU hidden size: 100 (XING), 1000 (TMALL)
        \item Dropout: 0.2 (XING), None (TMALL)
    \end{itemize}
    \item PNN
    \begin{itemize}
        \item Field embedding dimension: 10 (XING, TMALL)
        \item FC layer hidden size: 100 (XING), 300 (TMALL)
        \item Dropout: None (XING, TMALL)
    \end{itemize}
    \item ARNN
    \begin{itemize}
        \item Merging layer hidden size: 100 (XING), 1000 (TMALL)
        \item Dropout: None (XING, TMALL)
    \end{itemize}
\end{itemize}
As an optimizer, we used Adagrad\cite{adagrad} with different learning rates and weight decays for each dataset and model. It turned out that finding the right optimizer hyperparameters was crucial to the performance of the model.\footnote{For more details, please visit our Github repository. The repository will go public after the review process.}

\subsection{Evaluation}
\subsubsection{Measures}
We employed two ranking measures to evaluate the model performance: Recall@K and MRR@K. Let $N_{hits}$ be the number of times that a user chose an item from the recommendation list and $N_{recs}$ be the number of top-K recommendation attempts. Then, the measures can be calculated as follows:
\begin{equation*}
    Recall@K=\frac{N_{hits}}{N_{recs}} \qquad MRR@K=\frac{1}{N_{recs}}\sum_{n=1}^{N_{recs}}{\frac{1}{rank(n)}}
\end{equation*}
where $rank(n)$ denotes the rank of the chosen item within the $n$-th top-K recommendation list.
\subsubsection{Baselines}
We compared the ARNN with the following baselines:
\begin{itemize}
    \item Item-KNN\cite{item-based-cf-linden}: A simple yet powerful item-to-item approach that recommends items that are similar to the previous item based on cosine similarity.
    \item GRU4REC\cite{gru4rec}: A widely used RNN-based session model, as described in Section \ref{sess-rec-with-rnns}. Unlike ARNN, GRU4REC uses only the information from the item IDs.
    \item PNN\cite{product-based-nn}: A PNN context encoder, which is included to assess how well it captures high-order context information relevant for ranking.
\end{itemize}

\section{Results}

Table \ref{tab:evaluation} shows the Recall@20 and MRR@20 of the ARNN and the baselines for the two datasets.
\begin{table}[h!] \centering
    \caption{Evaluation results.}
    \vspace*{-0.3cm}
        \label{tab:evaluation}
        \begin{tabular}{lllll}
            \toprule
            & \multicolumn{2}{c}{XING} & \multicolumn{2}{c}{TMALL} \\
            \cmidrule{1-5}
            & Recall@20 & MRR@20 & Recall@20 & MRR@20\\
            \cmidrule{1-5}
            ItemKNN & 0.2235 & 0.0932 & 0.1541 & 0.0614 \\
            GRU4REC & 0.2597 & 0.1169 & 0.3789 & 0.1886 \\
            PNN & 0.1439 & 0.0507 & 0.1755 & 0.0694 \\
            \textbf{ARNN} & \textbf{0.3106} & \textbf{0.1252} & \textbf{0.3812} & \textbf{0.1953} \\
        \bottomrule
        \end{tabular}
    \vspace*{-0.3cm}
\end{table}

Evaluation of our model on XING provided us with a clear evidence that capturing user-contextual preference using a PNN is indeed helpful for RNN session models when rich user-side context is available. Although the performance of the PNN context encoder in itself was poor compared to GRU4REC, ARNN as a whole outperformed both ItemKNN and GRU4REC in both the Recall@20 and MRR@20 measures.

However, the performance of the ARNN was only \emph{marginally} better than the GRU4REC baseline in case of TMALL. Considering that only three user context variables were available for TMALL(user id, age, gender), our hypothesis is that the number of user context fields was not enough to augment new piece of information that could not be inferred by the GRU4REC. Thus, we recommend that ARNN should be used only when the number of user-side categorical fields is sufficiently large.

\section{Discussion and Conclusions}

In this study, we proposed an augmented RNN model that can easily boost an existing RNN session model by estimating high-order user-contextual preference using the PNN. Since PNN context encoders can handle \textit{arbitrary} user-side contextual information and build upon an existing pretrained RNN session model, we believe that deploying our model for real-world systems would be a handy solution that can improve the recommendation quality of a system without considerable effort.

However, one limitation of the ARNN is that it ignores the item-side contexts(e.g., item metadata, item content), unlike the usual factorization machines or PNNs used for CTR prediction. We omitted the item-side contexts because our intention was to measure the effect of incorporating \emph{pure} user-side contexts(e.g., age, location, login platform) with existing RNN session models using a FM-like approach. Therefore, designing an ARNN architecture that can also handle item-side contexts would be an interesting research topic for a future work.

\bibliographystyle{ACM-Reference-Format}
\bibliography{references}

\end{document}